\documentclass[12pt,english,floatfix,superscriptaddress,aps,prd,preprint,showkeys,nofootinbib]{revtex4}
\usepackage{amsmath}
\usepackage{amssymb}
\usepackage{amsbsy}
\usepackage{amsfonts}
\usepackage{amsopn}
\usepackage{amstext}
\usepackage{graphicx}
\usepackage[english]{babel}
\usepackage{color}
\usepackage{slashed}
\usepackage{esint}
\usepackage[dvips]{epsfig}
\usepackage[dvips]{graphicx}
\usepackage{float}
\usepackage{units}
\usepackage{textcomp}
\usepackage{wasysym}
\usepackage{hyperref}
\usepackage{slashed}

\newcommand{\trD}{\mathrm{tr}_{D}}
\newcommand{\Tr}{\mathrm{Tr}}
\newcommand{\Id}{\mathbf{1}}
\newcommand{\Kcal}{\mathcal{K}}
\newcommand{\Jcal}{\mathcal{J}}

\begin{document}

\title{Nonlocal four-fermion theory}

\author{F. M. Belchior}
\email{belchior@fisica.ufc.br}
\affiliation{Departamento de Física, Universidade Federal da Paraíba, Centro de Ciências Exatas e da Natureza, 58051-970, João Pessoa, Paraíba, Brazil}

\author{J. R. Nascimento}
\email{jroberto@fisica.ufpb.br}
\affiliation{Departamento de Física, Universidade Federal da Paraíba, Centro de Ciências Exatas e da Natureza, 58051-970, João Pessoa, Paraíba, Brazil}

\author{A. Yu. Petrov}
\email{petrov@fisica.ufpb.br}
\affiliation{Departamento de Física, Universidade Federal da Paraíba, Centro de Ciências Exatas e da Natureza, 58051-970, João Pessoa, Paraíba, Brazil}

 \author{P. J. Porfirio}
\email{pporfirio@fisica.ufpb.br}
\affiliation{Departamento de Física, Universidade Federal da Paraíba, Centro de Ciências Exatas e da Natureza, 58051-970, João Pessoa, Paraíba, Brazil}

\begin{abstract}
In this work, we formulate and analyze a nonlocal four-fermion theory in which the usual Dirac operator is deformed by an entire nonlocal form factor. After introducing an auxiliary scalar field, we derive the mean-field effective action and the corresponding gap equation for the dynamical mass. A central technical point of the analysis is that the nonlocal form factor is a matrix function of the Dirac operator, so the inverse propagator must be treated as an element of the closed algebra generated by the identity and $\slashed{p}$ operators, rather than as a purely scalar quantity. We obtain explicit expressions for the gap kernel for two representative choices of a form factor, $f_I(\slashed{\partial})=e^{-\slashed{\partial}/\Lambda}$ and $f_{II}(\slashed{\partial})=e^{-i\slashed{\partial}/\Lambda}$. Following the IR/UV matching method used in the recent Dirac-like nonlocal spinor theory \cite{Nascimento:2025ngc}, the momentum integral is split at an intermediate scale $M\ll \Omega\ll \Lambda$, expanded analytically in the infrared and ultraviolet regions, and compared with the usual local NJL/Gross-Neveu result. We show that the hyperbolic form factor enhances the gap integral and lowers the critical coupling, whereas the oscillatory form factor suppresses it and raises the critical coupling. The finite-temperature and finite-density extension is formulated through Matsubara sums and a corrected contour representation, with the local thermal gap equation recovered in the limit $\Lambda\to\infty$.
\end{abstract}

\maketitle

\section{Introduction}

Recently, nonlocal field theories have been studied as a framework to replace strictly point-like interactions by operators with an intrinsic length scale, usually through form factors such as $e^{\ell^2\Box}$ or, more generally, analytic functions of the d'Alembertian operator acting on fields (a general discussion of possible form factors is presented in \cite{BasiBeneito:2022wux}). The first motivations for this topic stem from attempts to soften ultraviolet behavior and eliminate divergences by dispersing  interactions in spacetime. Among these attempts, one highlights the Yukawa's proposal of nonlocal fields \cite{Yukawa:1950} and the axiomatic/constructive developments of nonlocal scalar QFT \cite{Efimov:1967}. Another origin of nonlocality comes from string theory, where effective actions in string field theory naturally generate infinitely many derivatives, and related nonlocal structures that appear in $p$-adic string models \cite{Brekke:1988,Frampton:1988,Eliezer:1989}. In this context, amplitudes and tachyon dynamics provide explicit examples of nonlocal kinetic operators and potentials \cite{Ghoshal:2000}. Moreover, the concept of nonlocality has also been employed as a gauge-invariant regulator, with explicit constructions of nonlocal regularizations for gauge theories \cite{Evens:1991}, and as a route to improved ultraviolet behavior in gauge and gravitational settings \cite{Tomboulis:1997}.

Particularly in the context of field theories, recent works have analyzed classes of nonlocal form factors that improve UV behavior while avoiding pathologies such as ghosts and acausality. In technical terms, equations with infinite derivatives can often be handled by localization methods that introduce auxiliary fields, clarifying degrees of freedom, and making computations manageable \cite{Calcagni:2008}. At the same time, the Cauchy problem is subtle in theories with infinitely many derivatives, where careful analyses show how well-posed evolution can still be defined in suitable analytic settings \cite{Barnaby:2008}. All these tools have enabled concrete applications ranging from cosmology, where $p$-adic inspired models yield distinctive inflationary dynamics \cite{Barnaby:2007}, to nonlocal gravity frameworks designed to resolve spacetime singularities and remain ghost-free \cite{Biswas:2012,Modesto:2012}. More systematic treatments of the structure of nonlocal and quasi-local theories \cite{Tomboulis:2015} and of ghost-free infinite-derivative QFT \cite{Buoninfante:2019} sharpen the criteria under which nonlocality improves UV behavior without introducing additional propagating instabilities, making nonlocal field theories a coherent and actively used effective description across particle physics and gravity.

In cosmology, recent proposals showed that nonlocal terms can drive late-time acceleration and mimic dark energy behavior \cite{Deser:2007jk,Koivisto:2008xfa}, and closely related nonlocal $F(R)$-type constructions were argued to unify inflationary and dark-energy eras within a single framework \cite{Nojiri:2007uq,Capozziello:2008gu}. More recent developments also investigated the phenomenology and conceptual status of nonlocal gravity models, including concrete cosmological predictions and tests \cite{Maggiore:2014sia,Belgacem:2017cqo}, as well as applications to causal G\"odel-type solutions \cite{Nascimento:2021bzb}. On the QFT side, nonlocality has been explored in the calculations of amplitudes \cite{Modesto:2021soh}. Other current works are \cite{Nascimento:2025ngc}, which examines formal consistency of a nonlocal spin-$\tfrac12$ dynamics, and \cite{Andrade:2024pce}, which analyzes nonlocal soliton/kink sectors in scalar theories.

This work is structured as follows: in section \ref{s2}, we review basic concepts of nonlocal field theories. In section \ref{s3}, we propose a nonlocal four-fermion model. In section \ref{s4}, we calculate the effective potential and the gap equation at zero temperature. We evaluate the resulting integral of the gap equation by employing cutoff and the IR/UV matching strategy used in the Dirac-like nonlocal theory of Ref.~\cite{Nascimento:2025ngc}. We also discuss nonlocal corrections and compare them with the local four-fermion model. In section \ref{s5}, we reexamine the theory at finite temperature. Finally, in section \ref{s6}, we present our final discussions and present future perspectives. 
 
\section{Nonlocal Field Theory}\label{s2}

In this section, we will present the basic idea of nonlocal field theory and the main issues it raises. Primarily, a nonlocal QFT is still a continuum field theory, but its dynamics cannot be captured by a Lagrangian involving only a finite number of derivatives. 
Instead, the action typically couples field values at separated spacetime points. Such nonlocality can appear in several ways. 
It may be an effective description obtained after integrating out heavy (or even massless) degrees of freedom, leading to operators such as $\phi\,\log(-\Box/\mu^2)\,\phi$ (such objects typically arise when loop corrections are calculated, see f.e. \cite{Kuzenko:1993tv,Petrov:2001pce,Buchbinder:2025jdy}, where such contributions are discussed within the supersymmetric field theories). It can also be introduced more fundamentally through UV-motivated infinite-derivative form factors, as in string-inspired models. Other mechanisms include long-range kernels, often described by fractional operators (see e.g. \cite{Almeida:2022nen}), and bilocal \cite{Iso:2000ew} or noncommutative \cite{Douglas:2001ba} interactions.

A convenient way to introduce nonlocality is through a kernel representation. For a scalar field, the most general Lorentz-invariant quadratic action can be written as follows
\begin{align}
S_2[\phi]=\frac{1}{2}\int d^4x\,d^4y\;\phi(x)\,K(x-y)\,\phi(y).
\label{eq1}
\end{align}
In momentum space, this becomes
\begin{align}
S_2=\frac{1}{2}\int\!\frac{d^4p}{(2\pi)^4}\;\tilde\phi(-p)\,\tilde K(p)\,\tilde\phi(p).\label{eq2}
\end{align}
When $\tilde K(p)$ depends only on $p^2$, one may write $\tilde K(p)=\mathcal{F}(-p^2)$ and equivalently express the action as
\begin{align}
S_2=\frac{1}{2}\int d^4x\;\phi\,\mathcal{F}(\Box)\,\phi,
\label{eq3}
\end{align}
where $\mathcal{F}(\Box)$ is understood as a pseudo-differential operator defined by its symbol $\mathcal{F}(-p^2)$. 
If $\mathcal{F}$ is analytic, it admits a formal expansion $\mathcal{F}(\Box)=\sum\limits_{n=0}^\infty c_n\Box^n$. 
In that case, the theory is nonlocal in the sense of involving infinitely many derivatives, but it can still behave as ``quasi-local'' at low energies if $\mathcal{F}$ is entire and introduces a characteristic scale $M$: for $E\ll M$ one can often use a local derivative expansion. By contrast, kernels with algebraic tails (for example $K\sim |x|^{-4-\alpha}$) represent genuinely long-range nonlocality that does not reduce to a controlled derivative series.

Another standard example of UV-softened field theory is described by the following action
\begin{align}
S=\int d^4x\left[\frac{1}{2}\,\phi\,e^{-\Box/M^2}(\Box-m^2)\phi
-\frac{\lambda}{4!}\phi^4\right],
\label{eq4}
\end{align}
and equation of motion
\begin{align}
e^{-\Box/M^2}(\Box-m^2)\phi+\frac{\lambda}{3!}\phi^3=0.
\label{eq5}
\end{align}
It is important to highlight that nonlocality can be present in the interaction sector. A simple example is a bilocal quartic coupling, namely
\begin{align}
S_{\rm int}
=-\frac{\lambda}{4!}\int d^4x\,d^4y\;f(x-y)\,\phi^2(x)\phi^2(y),
\label{eq6}
\end{align}
which ties bilinear field combinations at two different spacetime points. Another well-known source of nonlocal interactions comes from noncommutative star-products, where vertices pick up momentum-dependent phase factors, potentially leading to UV/IR mixing. For the free theory, the propagator takes the following schematic form
\begin{align}
\tilde G(p)=\frac{i\,e^{-p^2/M^2}}{p^2-m^2+i\varepsilon},
\label{eq7}
\end{align}
so high-momentum contributions are exponentially damped. 

The analytic structure of the form factor is where much of the physics lives. If the multiplier of $(p^2-m^2)$ is entire without zeros, with the exponential $e^{p^2/M^2}$ being the canonical example, so that no new poles are introduced and the spectrum is unchanged. If instead $\mathcal{F}(-p^2)$ has zeros, as when it factorizes into several quadratic factors, the propagator develops extra poles whose physical interpretation is controlled by the residues. Scanning the pole-and-cut structure of the two-point function is therefore the quickest consistency check.

Another analysis is regarded to causality that is discussed through the retarded Green's function given by
\begin{align}
\mathcal{F}(\Box)\,G_{\rm ret}(x)=\delta^{(4)}(x),
\label{eq8}
\end{align}
with $G_{\rm ret}(x)=0$ for $x^0<0$. For local hyperbolic operators, it is supported inside the light cone. Nonlocal form factors can spread propagation beyond it, generating spacetime tails. Microcausality is encoded in $[\phi(x),\phi(y)]=0$
for $(x-y)^2<0$ and its support is governed by the analyticity and pole-cut structure of the two-point function. Entire form factors without zeros can be consistent with perturbative unitarity, but do not automatically guarantee strict microcausality at all scales. Perturbative unitarity requires sensible pole residues and compatibility with the optical theorem. A related characteristic is related to spectral positivity, where in a local unitary QFT, the K\"{a}ll\'{e}n-Lehmann representation gives
\begin{align}
\langle 0|\phi(x)\phi(0)|0\rangle=\int_0^\infty d\mu^2\,\rho(\mu^2)\,\Delta_+(x;\mu^2),
\label{eq10}
\end{align}
with $\rho(\mu^2)\ge 0$ is the spectral density, and $\Delta_+(x;\mu^2)$ is the free Green function. In this case, nonlocality can modify the spectral density or alter the analyticity assumptions underlying dispersion relations.
Finally, the initial-value problem is governed by the zeros of $\mathcal{F}$ (or equivalently, the poles of the propagator) not by the formal order of the derivative expansion. If $\mathcal{F}$ shares its zeros with $(\Box-m^2)$, the theory propagates Klein-Gordon-like modes and does not require extra initial data. Additional zeros introduce new modes and a correspondingly more complicated Cauchy problem.

\section{Nonlocal four-fermion theory}\label{s3}

In this section, we will introduce a nonlocal fermionic theory with four-fermion interaction by generalizing the nonlocal spin-$\tfrac12$ theory of \cite{Nascimento:2025ngc}, where we consider a Dirac theory in which the usual $\slashed{\partial}$ is promoted to $\slashed{\partial}\,f(\slashed{\partial})$, with $f(z)$ an entire function of its argument $z$. Initially, let us make a brief review of this model, whose free Lagrangian reads
\begin{align}
\mathcal{L}_0=\bar\Psi_a\Big(i\,\slashed{\partial}\,f(\slashed{\partial})-m\Big)\Psi_a,
\label{eq11}
\end{align}
where $a=1,\dots,N$. We assume the $f(z)$ to be an analytic function which can be expanded in Taylor series as follows
\begin{align}
f(\slashed{\partial})=\sum_{n=0}^{\infty} b_n \left(\frac{\slashed{\partial}}{\Lambda}\right)^n,
\label{eq12}
\end{align}
where $b_0=1$ and $\Lambda$ is the nonlocality scale. Two suitable choices for $f$ can be (cf. \cite{Nascimento:2025ngc})
\begin{align}
f_{\rm I}(\slashed{\partial})=e^{-\slashed{\partial}/\Lambda},
\label{eq13}
\end{align}
and
\begin{align}
 f_{\rm II}(\slashed{\partial})=e^{-i\,\slashed{\partial}/\Lambda}.
\label{eq14}   
\end{align}
Besides, from \eqref{eq11}, the momentum-space Feynman propagator is
\begin{align}
S_F(p)=\frac{i}{\slashed{p}\,f(\slashed{p})-m}=i\,\frac{\slashed{p}\,f(\slashed{p})+m}{p^2 f^2(\slashed{p})-m^2}.
\label{eq15}
\end{align}
A key consistency requirement is that $f$ be entire and chosen so that no new zeros are introduced in the physical sheet beyond the standard mass shell; then the spectrum is not enlarged by additional poles. The simplest nonlocal four-fermion theory keeps the interaction local but uses the nonlocal propagator \eqref{eq15}. For $N$ flavors, a convenient large-$N$ normalization is
\begin{align}
\mathcal{L}_{4\psi}=\frac{G}{2N}\Big(\bar\Psi_a\Psi_a\Big)^2,
\label{eq16}
\end{align}
which preserves global $U(1)$ symmetry. The index $a$ takes values from 1 to $N$. Here and further, the sum over repeating isotopic indices is assumed. An interaction invariant under chiral transformations (the chiral symmetry is, as usual, possible only for $m=0$) reads
\begin{align}
\mathcal{L}_{4\psi}^{\chi}=\frac{G}{2N}\Big[\big(\bar\Psi_a\Psi_a\big)^2+\big(\bar\Psi_a i\gamma^5\Psi_a\big)^2\Big],
\label{eq17}
\end{align}
and one may also consider vector/axial/tensor channels, subject to the usual Fierz relations. The full model we explore in this study is chosen to be
\begin{align}
\mathcal{L}=\bar\Psi_a\Big(i\,\slashed{\partial}\,f(\slashed{\partial})-m\Big)\Psi_a+\frac{G}{2N}\Big(\bar\Psi_a\Psi_a\Big)^2,
\label{eq18}
\end{align}
with Euler-Lagrange equation
\begin{align}
\Big(i\,\slashed{\partial}\,f(\slashed{\partial})-m\Big)\Psi_a
+\frac{G}{N}\Big(\bar\Psi_b\Psi_b\Big)\Psi_a=0.
\label{eq19}
\end{align}

To describe a model with $N$ fields (in this case, the spinor ones), one can use the $1/N$ expansion framework (for a review on it, see f.e. \cite{Coleman:1980nk,Moshe:2003xn}; examples of applying this machinery can be found also in \cite{Koures:1991zu,Girotti:2001ku}). To do it, we start with introducing a real auxiliary Lagrange multiplier scalar field $\sigma(x)$, so that
\begin{align}
\mathcal{L}_{\mathrm{HS}}=\bar\Psi_a\Big(i\,\slashed{\partial}\,f(\slashed{\partial})-m-\sigma\Big)\Psi_a-\frac{N}{2G}\,\sigma^2.
\label{eq20}
\end{align}
This Lagrangian multiplier, as usual within this framework, serves to eliminate the non-renormalizable four-fermion interaction.
Integrating out $\sigma$, one reproduces \eqref{eq16}. 

Here, it is important to explain a nomenclature point for the physical interpretation of the present construction. In four spacetime dimensions, a local scalar-channel four-fermion theory used to describe chiral symmetry breaking and fermion condensation is most commonly called the Nambu-Jona-Lasinio (NJL) model, whereas the Gross-Neveu (GN) name is usually associated with the asymptotically free model in two dimensions and with related lower-dimensional large-$N$ generalizations. Since our formulation is four-dimensional but uses the same large-$N$ auxiliary-field logic that also appears in GN-type models, we shall refer to it as a nonlocal Nambu-Jona-Lasinio/Gross-Neveu (NJL/GN) model. This terminology emphasizes that the four-dimensional limit should be compared primarily with the NJL literature, while the mean-field and large-$N$ structure remain naturally connected with the GN tradition \cite{Nambu:1961tp,Nambu:1961fr,Gross:1974jv,Klevansky:1992qe,Hatsuda:1994pi}.

Although we introduced the mass in our study from the very beginning in (\ref{eq11}), it should be noted that in principle, it can be generated from the action (\ref{eq20}) considered at $m=0$ (the similar approach has been used, e.g. in \cite{Koures:1991zu,Girotti:2001ku}). To do it, we assign to the massless $\sigma$ field an expectation value $<\sigma>=m$, make a shift $\sigma\to\sigma+m$ (we note, however, that such a shift breaks a chiral symmetry), and as a result we arrive at the massive Lagrangian (\ref{eq20}) plus an irrelevant term linear in $\sigma$ which yields trivial contributions within our study.

This mechanism should be viewed as the standard fermion-condensation mechanism of NJL/GN-type theories. The auxiliary field is not introduced as a fundamental scalar degree of freedom; rather, its expectation value parametrizes the composite bilinear condensate $\langle\bar\Psi\Psi\rangle$, which plays the role of an order parameter for dynamical mass generation. In four dimensions this idea goes back to the original NJL construction, where the gap equation is the relativistic analogue of the BCS self-consistency equation, and it has since become a standard effective description of chiral symmetry breaking in strong-interaction phenomenology \cite{Nambu:1961tp,Nambu:1961fr,Klevansky:1992qe,Hatsuda:1994pi}. The GN model and its large-$N$ extensions provide a complementary lower-dimensional laboratory in which the same relation between a four-fermion attraction, a nonzero bilinear condensate, and a dynamically generated fermion mass can be analyzed with high analytic control \cite{Gross:1974jv,Coleman:1980nk,Moshe:2003xn}.

\section{Effective potential and matrix-valued gap equation}\label{s4}

In this section, we will derive the effective potential directly from the path integral. This step is useful because it fixes the relative sign between the auxiliary-field term and the fermionic determinant, and it also makes the large-$N$ counting explicit.

Before entering the explicit calculation, let us highlight the physical motivation for computing the effective potential and the gap equation in this context. The gap equation determines the stationary values of the dynamical mass, while the effective potential contains more information,  telling us which stationary solution is thermodynamically preferred, whether the trivial solution $M=0$ is stable or unstable, and how the nonlocal scale $\Lambda$ reshapes the energy landscape associated with fermion condensation. In this sense, $V_{\mathrm{eff}}$ is the natural object for discussing spontaneous mass generation, metastability, possible first-order transitions, and the comparison between different choices of entire form factor. The derivation below is both a technical route to the gap equation and the basis for interpreting the nonlocal theory as a deformation of the usual Gross-Neveu theory.

At this point, it is worth highlighting the concepts of fermion condensation and dynamical mass generation in this context. A nonzero stationary value of $M$ corresponds, through the Hubbard-Stratonovich relation  (for a general review on this approach, see e.g. \cite{Kleinert:2011rb}), to a nonzero scalar bilinear condensate. This point connects the present nonlocal calculation with the extensive literature on NJL model studying its various issues, such as chiral condensates, effective potentials, and medium-dependent order parameters \cite{Klevansky:1992qe,Hatsuda:1994pi,Buballa:2003qv}. The novelty of our analysis is not the existence of the condensate itself, but the way in which the matrix-valued nonlocal Dirac operator changes the fermionic determinant that drives the condensation.

The Hubbard-Stratonovich-like form of the model is
\begin{equation}
\mathcal{L}_{\mathrm{HS}}=\bar\Psi_a\Big(i\slashed{\partial}f(\slashed{\partial})-m-\sigma\Big)\Psi_a-\frac{N}{2G}\sigma^2 .
\label{hs-fixed}
\end{equation}
The normalization in Eq.~\eqref{hs-fixed} is chosen so that integrating over the real auxiliary field gives back the original interaction $G(\bar\Psi_a\Psi_a)^2/(2N)$. Indeed, at each spacetime point,
\begin{equation}
\int d\sigma\,\exp\left\{i\left[-\frac{N}{2G}\sigma^2-\sigma\,\bar\Psi_a\Psi_a\right]\right\}
\propto
\exp\left\{i\frac{G}{2N}\left(\bar\Psi_a\Psi_a\right)^2\right\},
\label{hs-identity}
\end{equation}
where flavor indices are summed. Therefore the auxiliary-field representation is exactly equivalent to the four-fermion theory before the saddle-point approximation is taken.

Let us define the Dirac operator in the background $\sigma(x)$ as
\begin{equation}
D_\sigma \equiv i\slashed{\partial}f(\slashed{\partial})-m-\sigma(x).
\label{Dsigma}
\end{equation}
The generating functional in the presence of Grassmann sources is
\begin{align}
Z[\bar\eta,\eta]=&\int \mathcal{D}\sigma\,\mathcal{D}\bar\Psi\,\mathcal{D}\Psi\,
\exp\Bigg\{i\int d^4x\left[-\frac{N}{2G}\sigma^2
+\bar\Psi_aD_\sigma\Psi_a+\bar\eta_a\Psi_a+\bar\Psi_a\eta_a\right]\Bigg\} .
\label{Zsources}
\end{align}
For fixed $\sigma$, the fermionic part is Gaussian. Completing the square,
$\Psi_a\rightarrow \Psi_a-D_\sigma^{-1}\eta_a$ and $\bar\Psi_a\rightarrow \bar\Psi_a-\bar\eta_aD_\sigma^{-1}$, and using the invariance of the Grassmann measure, one obtains
\begin{align}
\int \mathcal{D}\bar\Psi\,\mathcal{D}\Psi\,
\exp\left\{i\int d^4x\left(\bar\Psi_aD_\sigma\Psi_a+\bar\eta_a\Psi_a+\bar\Psi_a\eta_a\right)\right\}
\nonumber=\\\left[\det D_\sigma\right]^N
\exp\left\{-i\int d^4x\,d^4y\;\bar\eta_a(x)D_\sigma^{-1}(x,y)\eta_a(y)\right\} .
\label{fermion-gaussian}
\end{align}
Thus
\begin{equation}
Z[\bar\eta,\eta]=\int \mathcal{D}\sigma\,
\exp\left\{iS_{\mathrm{eff}}^{M}[\sigma]
-i\int d^4x\,d^4y\;\bar\eta_a(x)D_\sigma^{-1}(x,y)\eta_a(y)\right\},
\label{eq:Zeff}
\end{equation}
with the Minkowski effective action
\begin{equation}
S_{\mathrm{eff}}^{M}[\sigma]=
-N\int d^4x\,\frac{\sigma^2}{2G}
-iN\,\Tr\ln D_\sigma .
\label{SeffM}
\end{equation}
Here $\Tr$ denotes the functional trace over spacetime and Dirac indices, while the factor $N$ comes from the flavor trace. Equation~\eqref{SeffM} is the starting point of the leading large-$N$ approximation: fluctuations of $\sigma$ are suppressed by powers of $1/N$, while the saddle contribution scales as $N$.

We now take a translationally invariant saddle $\sigma(x)=\bar\sigma$ and $M=m+\bar\sigma$. After Wick rotation, and dropping field-independent constants, the Euclidean effective action becomes $\Gamma_E=N V_4 V_{\mathrm{eff}}(M)$, where $V_4$ is the Euclidean four-volume. Because the Dirac-like form factor is matrix valued and does not need be an even function of $\slashed p$, the unsymmetrized Euclidean determinant can contain a phase. Following the Euclidean prescription used for the nonlocal Dirac-like spinor theory, we define the real mean-field potential from the paired determinant given by
\begin{equation}
\ln{\rm det}_D D_E(p)\;\longrightarrow\;
\frac{1}{2}\ln{\rm det}_D\left[D_+(p)D_-(p)\right],
\label{paired-det}
\end{equation}
with $D_\pm(p)=M\pm i\slashed p\,f(\slashed p)$. Since $f(\slashed p)$ commutes with $\slashed p$, one has
\begin{equation}
D_+(p)D_-(p)=M^2+p^2f^2(\slashed p).
\label{eq:squared-dirac}
\end{equation}
Thus, the effective potential per fermion flavor is
\begin{equation}
V_{\mathrm{eff}}(M)=\frac{(M-m)^2}{2G}-\frac{1}{2}\int\!\frac{d^4p}{(2\pi)^4}\ln{\rm det}_D\left[p^2f^2(\slashed p)+M^2\right] .
\label{veff}
\end{equation}

This expression is one of the central results of the section since it shows that the mean-field potential is composed of a classical auxiliary-field contribution, $(M-m)^2/(2G)$, and a quantum fermionic determinant modified by the nonlocal operator. The important point is that the nonlocality does not enter as a simple multiplicative correction to the local determinant. Instead, it changes the matrix structure inside the logarithm through $f^2(\slashed p)$, and only after the Dirac determinant is evaluated can one identify the scalar function that controls the potential. This is why the matrix-valued nature of the form factor must be kept until the end of the calculation.

To proceed further, it is useful to write Eq.~\eqref{veff} in a form where the Dirac determinant has already been reduced to a scalar logarithm. Using the decomposition in Eq.~\eqref{decomp} below, one has
\begin{equation}
{\rm det}_D\left[A(p)\Id+B(p)\slashed p\right]=\left[A^2(p)-B^2(p)p^2\right]^2 .
\label{detAB}
\end{equation}
Thus, after angular integration, the zero-temperature potential can be written as
\begin{equation}
V_{\mathrm{eff}}(M)=\frac{(M-m)^2}{2G}-\frac{1}{8\pi^2}\int_0^{\Lambda_{\mathrm{UV}}}dp\,p^3
\ln\left[A^2(p)-B^2(p)p^2\right] .
\label{veffAB}
\end{equation}
The additive constant obtained by setting $M=0$ has no physical effect on the gap equation. If the subtracted potential is defined literally as $\Delta V_{\mathrm{eff}}(M)=V_{\mathrm{eff}}(M)-V_{\mathrm{eff}}(0)$, the classical contribution contains the constant $-m^2/(2G)$. We therefore write
\begin{equation}
\Delta V_{\mathrm{eff}}(M)\equiv V_{\mathrm{eff}}(M)-V_{\mathrm{eff}}(0)
=\frac{(M-m)^2-m^2}{2G}-\frac{1}{8\pi^2}\,\mathcal{I}_f(M).
\label{deltaV}
\end{equation}
This extra constant is irrelevant for the gap equation, but keeping it makes the definition of the subtracted potential unambiguous.
Here, we have
\begin{equation}
\mathcal{I}_f(M)=\int_0^{\Lambda_{\mathrm{UV}}}dp\,p^3\ln\left[\frac{A^2(p)-B^2(p)p^2}{p^4}\right].
\label{Ifdef}
\end{equation}
For the local theory, this gives the following closed expression
\begin{align}
\mathcal{I}_{\mathrm{loc}}(M;R)&=2\int_0^R dp\,p^3\ln\left(1+\frac{M^2}{p^2}\right)\nonumber\\
&=\frac{R^4}{2}\ln\left(1+\frac{M^2}{R^2}\right)+\frac{M^2R^2}{2}
-\frac{M^4}{2}\ln\left(1+\frac{R^2}{M^2}\right) .
\label{Ieffloc}
\end{align}
The factors of $1/2$ in the second line are important: with this normalization the derivative of the logarithmic integral is exactly the one needed by the local gap equation.

The derivative obeys the useful relation:
\begin{equation}
\frac{d\mathcal{I}_{\mathrm{loc}}(M;R)}{dM}=4M\,\mathcal{J}_{\mathrm{loc}}(M;R),
\qquad
\mathcal{J}_{\mathrm{loc}}(M;R)=\frac12\left[R^2-M^2\ln\left(1+\frac{R^2}{M^2}\right)\right],
\label{Ieffloc-check}
\end{equation}
so that differentiating the local potential reproduces the local gap equation exactly. Therefore, one has
\begin{equation}
\Delta V_{\mathrm{loc}}(M)=\frac{(M-m)^2-m^2}{2G}
-\frac{1}{8\pi^2}\mathcal{I}_{\mathrm{loc}}(M;\Lambda_{\mathrm{UV}}).
\label{Vloc-final}
\end{equation}

Substituting Eq.~\eqref{Ieffloc} into Eq.~\eqref{Vloc-final}, the local cutoff potential can be written explicitly as
\begin{align}
\Delta V_{\mathrm{loc}}(M)
&=\frac{(M-m)^2-m^2}{2G}
-\frac{1}{8\pi^2}\left[
\frac{\Lambda_{\mathrm{UV}}^4}{2}\ln\left(1+\frac{M^2}{\Lambda_{\mathrm{UV}}^2}\right)
+\frac{M^2\Lambda_{\mathrm{UV}}^2}{2}
-\frac{M^4}{2}\ln\left(1+\frac{\Lambda_{\mathrm{UV}}^2}{M^2}\right)
\right].
\label{Vloc-explicit-CW}
\end{align}
This formula makes the relation with the usual Coleman-Weinberg mechanism transparent. In the chiral limit, $m=0$, and for $M\ll\Lambda_{\mathrm{UV}}$, one obtains
\begin{align}
\Delta V_{\mathrm{loc}}(M)
&=\frac{M^2}{2}\left(\frac{1}{G}-\frac{\Lambda_{\mathrm{UV}}^2}{4\pi^2}\right)
+\frac{M^4}{16\pi^2}\ln\left(\frac{\Lambda_{\mathrm{UV}}^2}{M^2}\right)
+\frac{M^4}{32\pi^2}
+\mathcal{O}\left(\frac{M^6}{\Lambda_{\mathrm{UV}}^2}\right).
\label{Vloc-CW-expansion}
\end{align}
Equivalently, by introducing the local critical coupling
\begin{equation}
\frac{1}{G_c^{\mathrm{loc}}}=\frac{\Lambda_{\mathrm{UV}}^2}{4\pi^2},
\label{Gc-local-CW}
\end{equation}
Eq.~\eqref{Vloc-CW-expansion} becomes
\begin{equation}
\Delta V_{\mathrm{loc}}(M)
=\frac{1}{2}\left(\frac{1}{G}-\frac{1}{G_c^{\mathrm{loc}}}\right)M^2
+\frac{M^4}{16\pi^2}\ln\left(\frac{\Lambda_{\mathrm{UV}}^2}{M^2}\right)
+\frac{M^4}{32\pi^2}+\cdots .
\label{Vloc-CW-critical}
\end{equation}
Thus, after the cutoff-dependent quadratic term is isolated, the leading nontrivial quantum correction is the characteristic logarithmic term $-M^4\ln(M^2/\Lambda_{\mathrm{UV}}^2)/(16\pi^2)$, together with additive polynomial terms that can be absorbed into the usual effective-theory parameters. This is the four-fermion, composite-order-parameter analogue of the Coleman-Weinberg effective potential. The nonlocal expressions in Eqs.~\eqref{VI-final} and \eqref{VII-final} must reduce to this result when $f(\slashed p)\rightarrow 1$, or equivalently when $\Lambda\rightarrow\infty$ at fixed $p$ and $M$.

For the two Dirac-like form factors considered here, the subtracted logarithmic integrals are
\begin{align}
\mathcal{I}_{I}(M)&=\int_0^{\Lambda_{\mathrm{UV}}}dp\,p^3
\ln\left[1+\frac{2M^2}{p^2}\cosh\left(\frac{2p}{\Lambda}\right)+\frac{M^4}{p^4}\right],
\label{IeffI}\\
\mathcal{I}_{II}(M)&=\int_0^{\Lambda_{\mathrm{UV}}}dp\,p^3
\ln\left[1+\frac{2M^2}{p^2}\cos\left(\frac{2p}{\Lambda}\right)+\frac{M^4}{p^4}\right].
\label{IeffII}
\end{align}
Thus the final zero-temperature potentials are
\begin{align}
\Delta V_I(M)&=\frac{(M-m)^2-m^2}{2G}-\frac{1}{8\pi^2}\mathcal{I}_I(M),
\label{VI-final}\\
\Delta V_{II}(M)&=\frac{(M-m)^2-m^2}{2G}-\frac{1}{8\pi^2}\mathcal{I}_{II}(M).
\label{VII-final}
\end{align}

Equivalently, without using the shorthand notation for the logarithmic integrals, the two nonlocal mean-field potentials are
\begin{align}
\Delta V_I(M)
&=\frac{(M-m)^2-m^2}{2G}
-\frac{1}{8\pi^2}\int_0^{\Lambda_{\mathrm{UV}}}dp\,p^3
\ln\left[1+\frac{2M^2}{p^2}\cosh\left(\frac{2p}{\Lambda}\right)+\frac{M^4}{p^4}\right],
\label{VI-explicit-nonlocal}\\
\Delta V_{II}(M)
&=\frac{(M-m)^2-m^2}{2G}
-\frac{1}{8\pi^2}\int_0^{\Lambda_{\mathrm{UV}}}dp\,p^3
\ln\left[1+\frac{2M^2}{p^2}\cos\left(\frac{2p}{\Lambda}\right)+\frac{M^4}{p^4}\right].
\label{VII-explicit-nonlocal}
\end{align}
These equations are the exact cutoff representations of the nonlocal effective potentials at leading order in the large-$N$ expansion. In contrast with the local case, the logarithms contain the full non-polynomial dependence on the nonlocality scale $\Lambda$. For this reason the nonlocal potentials are most transparently written as the one-dimensional momentum integrals above, while their analytic content is extracted by the IR/UV expansion described below.

The local Coleman-Weinberg analogue is recovered directly from Eqs.~\eqref{VI-explicit-nonlocal} and \eqref{VII-explicit-nonlocal}. Introducing
\begin{equation}
C_\eta(x)=
\begin{cases}
\cosh x, & \eta=+1,\\
\cos x, & \eta=-1,
\end{cases}
\end{equation}
we can write both kernels as
\begin{equation}
\ln\left[1+\frac{2M^2}{p^2}C_\eta\left(\frac{2p}{\Lambda}\right)+\frac{M^4}{p^4}\right].
\end{equation}
For fixed $p$ and $M$, the large-$\Lambda$ expansion gives
\begin{align}
C_\eta\left(\frac{2p}{\Lambda}\right)&=1+\eta\frac{2p^2}{\Lambda^2}+\mathcal{O}\left(\frac{p^4}{\Lambda^4}\right),\\\ln\left[1+\frac{2M^2}{p^2}C_\eta\left(\frac{2p}{\Lambda}\right)+\frac{M^4}{p^4}\right]&=2\ln\left(1+\frac{M^2}{p^2}\right)+\eta\frac{4M^2p^4}{\Lambda^2(p^2+M^2)^2}+\mathcal{O}\left(\frac{1}{\Lambda^4}\right).
\label{nonlocal-potential-local-expansion}
\end{align}
Consequently, we find
\begin{equation}
\Delta V_\eta(M)=\Delta V_{\mathrm{loc}}(M)
-\eta\frac{M^2}{2\pi^2\Lambda^2}\,\mathcal{B}(M,\Lambda_{\mathrm{UV}})
+\mathcal{O}\left(\frac{1}{\Lambda^4}\right),
\label{Veta-local-limit}
\end{equation}
where
\begin{equation}
\mathcal{B}(M,R)=\int_0^R dp\,\frac{p^7}{(p^2+M^2)^2}=\frac{R^4}{4}-M^2R^2+\frac{3M^4}{2}\ln\left(1+\frac{R^2}{M^2}\right)-\frac{M^4R^2}{2(R^2+M^2)} .
\label{Bfunc-nonlocal-limit}
\end{equation}
Thus $\Delta V_\eta(M)\to\Delta V_{\mathrm{loc}}(M)$ when $\Lambda\to\infty$. The first correction has opposite signs for the two kernels: the hyperbolic form factor $(\eta=+1)$ lowers the potential relative to the local theory, while the oscillatory form factor $(\eta=-1)$ raises it. This gives the potential-level explanation of why the first kernel favors condensation and the second one suppresses it.

Equations~\eqref{Vloc-final}, \eqref{VI-final}, and \eqref{VII-final} are the final cutoff effective potentials used in the numerical analysis.

It is useful to highlight the final result in a compact form. For the local theory and for two nonlocal form factors considered in this work, the zero-temperature effective potentials are summarized by
\begin{align}
\Delta V_{\mathrm{loc}}(M)=\frac{(M-m)^2-m^2}{2G}
-\frac{1}{8\pi^2}\mathcal{I}_{\mathrm{loc}}(M;\Lambda_{\mathrm{UV}}),
\label{Vloc-highlight}\\
\Delta V_I(M)=\frac{(M-m)^2-m^2}{2G}-\frac{1}{8\pi^2}\mathcal{I}_I(M),
\label{VI-highlight}\\
\Delta V_{II}(M)=\frac{(M-m)^2-m^2}{2G}-\frac{1}{8\pi^2}\mathcal{I}_{II}(M).
\label{VII-highlight}
\end{align}
Together with Eqs.~\eqref{Ieffloc}, \eqref{IeffI}, and \eqref{IeffII}, these equations represent the main outcome of the calculation. They tell us how the same classical term competes with three different fermionic contributions: the local determinant, the hyperbolic nonlocal determinant, and the oscillatory nonlocal determinant. Therefore, the change in the vacuum structure is entirely encoded in the logarithmic integrals $\mathcal{I}_I$ and $\mathcal{I}_{II}$, whose signs and magnitudes determine whether the nonlocal operator favors or suppresses the fermion condensation.

We can evaluate the nonlocal logarithmic integrals analytically in the same spirit as Ref.~\cite{Nascimento:2025ngc}, where we can introduce an intermediate scale $\Omega$ satisfying $M\ll\Omega\ll\Lambda$, expand the integrand in powers of $p/\Lambda$ in the IR region, and expand it in powers of $M/p$ in the UV region. Denoting $\eta=+1$ for $f_I$ and $\eta=-1$ for $f_{II}$, the IR logarithm is
\begin{equation}
\ln\left[\frac{D_\eta(p;M)}{p^4}\right]=2\ln\left(1+\frac{M^2}{p^2}\right)
+\eta\frac{4M^2p^4}{\Lambda^2(p^2+M^2)^2}+\mathcal{O}\left(\frac{p^6}{\Lambda^4}\right),
\label{IRlogV}
\end{equation}
where $D_I=p^4+2M^2p^2\cosh(2p/\Lambda)+M^4$ and $D_{II}=p^4+2M^2p^2\cos(2p/\Lambda)+M^4$. Hence, we have
\begin{equation}
\mathcal{I}^{\mathrm{IR}}_\eta(M,\Omega)=\mathcal{I}_{\mathrm{loc}}(M;\Omega)+\eta\frac{4M^2}{\Lambda^2}\mathcal{B}(M,\Omega)
+\mathcal{O}\left(\frac{\Omega^8}{\Lambda^4}\right),
\label{IIRV}
\end{equation}
with
\begin{align}
\mathcal{B}(M,\Omega)&=\int_0^\Omega dp\,\frac{p^7}{(p^2+M^2)^2}
\nonumber\\
&=\frac{\Omega^4}{4}-M^2\Omega^2+\frac{3M^4}{2}
\ln\left(1+\frac{\Omega^2}{M^2}\right)
-\frac{M^4\Omega^2}{2(\Omega^2+M^2)} .
\label{BfuncV}
\end{align}
In the UV region, $p\gg M$, the logarithm becomes
\begin{align}
\ln\left[\frac{D_I(p;M)}{p^4}\right]
&=\frac{2M^2}{p^2}\cosh\left(\frac{2p}{\Lambda}\right)+\mathcal{O}\left(\frac{M^4}{p^4}\right),
\label{eq:UVlogI}\\
\ln\left[\frac{D_{II}(p;M)}{p^4}\right]
&=\frac{2M^2}{p^2}\cos\left(\frac{2p}{\Lambda}\right)+\mathcal{O}\left(\frac{M^4}{p^4}\right).
\label{UVlogII}
\end{align}
Therefore the UV contributions are
\begin{align}
\mathcal{I}^{\mathrm{UV}}_I(M,
\Omega,\Lambda_{\mathrm{UV}})
&=2M^2\left[
\frac{\Lambda p}{2}\sinh\left(\frac{2p}{\Lambda}\right)
-\frac{\Lambda^2}{4}\cosh\left(\frac{2p}{\Lambda}\right)
\right]_{\Omega}^{\Lambda_{\mathrm{UV}}}
+\mathcal{O}(M^4),
\label{IUVV-I}\\
\mathcal{I}^{\mathrm{UV}}_{II}(M,
\Omega,\Lambda_{\mathrm{UV}})
&=2M^2\left[
\frac{\Lambda p}{2}\sin\left(\frac{2p}{\Lambda}\right)
+\frac{\Lambda^2}{4}\cos\left(\frac{2p}{\Lambda}\right)
\right]_{\Omega}^{\Lambda_{\mathrm{UV}}}
+\mathcal{O}(M^4).
\label{IUVV-II}
\end{align}
The IR/UV approximation to the effective potential is then
\begin{align}
\Delta V_I^{\mathrm{IR/UV}}(M)&=\frac{(M-m)^2-m^2}{2G}
-\frac{1}{8\pi^2}\left[\mathcal{I}^{\mathrm{IR}}_{+}(M,\Omega)+
\mathcal{I}^{\mathrm{UV}}_I(M,
\Omega,\Lambda_{\mathrm{UV}})\right],
\label{VIRUV-I}\\
\Delta V_{II}^{\mathrm{IR/UV}}(M)&=\frac{(M-m)^2-m^2}{2G}
-\frac{1}{8\pi^2}\left[\mathcal{I}^{\mathrm{IR}}_{-}(M,\Omega)+
\mathcal{I}^{\mathrm{UV}}_{II}(M,
\Omega,\Lambda_{\mathrm{UV}})\right].
\label{VIRUV-II}
\end{align}
Differentiating Eqs.~\eqref{VIRUV-I} and \eqref{VIRUV-II} with respect to $M$ reproduces the IR/UV gap equations displayed below, up to the same truncation order. The local split is recovered by replacing the UV form factors by unity,
\begin{equation}
\mathcal{I}^{\mathrm{split}}_{\mathrm{loc}}(M)=
\mathcal{I}_{\mathrm{loc}}(M;\Omega)+M^2(\Lambda_{\mathrm{UV}}^2-\Omega^2)
-M^4\ln\left(\frac{\Lambda_{\mathrm{UV}}}{\Omega}\right)
+\mathcal{O}\left(\frac{M^6}{\Omega^2}\right),
\label{Iefflocsplit}
\end{equation}
which is the potential-level analog of the split gap integral. The explicit logarithmic term is important: without it the differentiated split potential does not reproduce the local gap equation at the same order.

The interpretation of Eqs.~\eqref{VIRUV-I} and \eqref{VIRUV-II} is also worth emphasizing. The IR part controls the smooth deformation of the local low-energy theory, whereas the UV part measures how strongly the nonlocal form factor weights modes near the intrinsic scale $\Lambda$. For the hyperbolic form factor the UV contribution grows, lowering the energy cost of forming a condensate. For the oscillatory form factor the UV contribution is partially cancelled by sign-changing trigonometric behavior, making condensation less efficient. Thus the effective potential provides a direct energetic explanation for the shift of the critical coupling found later from the gap equation.

The factor $i$ multiplying the Euclidean kinetic term is essential: since $(i\slashed p)^2=-p^2\Id$, the local limit gives $\frac12\ln\det_D(p^2+M^2)=2\ln(p^2+M^2)$, namely the usual Dirac determinant contribution.

The mean-field equation is obtained by imposing stationarity with respect to $M$, namely
\begin{equation}
0=\frac{\partial V_{\mathrm{eff}}}{\partial M}
=\frac{M-m}{G}
-M\,\trD\int\!\frac{d^4p}{(2\pi)^4}
\left[p^2f^2(\slashed p)+M^2\right]^{-1} .
\label{stationarity}
\end{equation}
Therefore
\begin{equation}
\frac{M-m}{G}=M\,\trD\int\!\frac{d^4p}{(2\pi)^4}
\left[p^2f^2(\slashed p)+M^2\right]^{-1} .
\label{gap-trace}
\end{equation}
This equation is written per flavor. If one keeps the full effective potential instead of $V_{\mathrm{eff}}/N$, both sides of the effective action carry the same overall factor $N$, and Eq.~\eqref{gap-trace} is unchanged. Notice also that Eqs.~\eqref{veff}-\eqref{gap-trace} do not assume that the form factor is a scalar function; the matrix nature of $f(\slashed p)$ must still be handled before the Dirac trace is evaluated.

The nonlocal form factor is not a scalar in spinor space. We therefore use the Euclidean Clifford algebra $\{\gamma_\mu,\gamma_\nu\}=2\delta_{\mu\nu}$, so that $\slashed p^2=p^2\Id$ and $(i\slashed p)^2=-p^2\Id$. With the form factor evaluated as a matrix function of $\slashed p$, we decompose
\begin{equation}
f^2(\slashed p)=a_1(p)\Id+a_2(p)\slashed p,
\label{decomp}
\end{equation}
where $p=\sqrt{p^2}$ and $a_1,a_2$ are scalar functions. The scalar denominator entering the traced propagator is
\begin{equation}
\mathcal{D}(p)=p^2f^2(\slashed p)+M^2=A(p)\Id+B(p)\slashed p,
\label{DAB}
\end{equation}
with
\begin{equation}
A(p)=p^2a_1(p)+M^2    
\end{equation}
and
\begin{align}
B(p)=p^2a_2(p)    
\end{align}
Using $\slashed p^2=p^2\Id$, one obtains
\begin{equation}
\mathcal{D}^{-1}(p)=\frac{A(p)\Id-B(p)\slashed p}{A^2(p)-B^2(p)p^2} .
\label{matrix-inverse}
\end{equation}
Only the identity part contributes to the Dirac trace. Thus the gap equation becomes
\begin{equation}
\frac{M-m}{G}=4M\int\!\frac{d^4p}{(2\pi)^4}\,\Kcal_f(p;M,\Lambda),
\label{gap-general}
\end{equation}
with
\begin{equation}
\Kcal_f(p;M,\Lambda)=
\frac{p^2a_1(p)+M^2}{\left[p^2a_1(p)+M^2\right]^2-p^6a_2^2(p)} .
\label{kernel-general}
\end{equation}
The angular integration gives the one-dimensional form
\begin{equation}
\frac{M-m}{G}=\frac{M}{2\pi^2}\,\Jcal_f(M,\Lambda,\Lambda_{\mathrm{UV}}),
\label{Jdef}
\end{equation}
with
\begin{align}
\Jcal_f\equiv\int_0^{\Lambda_{\mathrm{UV}}}\!dp\,p^3\Kcal_f(p;M,\Lambda)    
\end{align}
The local limit $f\to1$ gives
\begin{equation}
\Kcal_{\mathrm{loc}}(p;M)=\frac{1}{p^2+M^2},
\quad
\Jcal_{\mathrm{loc}}=\frac{1}{2}\left[\Lambda_{\mathrm{UV}}^2-M^2\ln\left(1+\frac{\Lambda_{\mathrm{UV}}^2}{M^2}\right)\right].
\label{localJ}
\end{equation}
Equation~\eqref{localJ} is the standard cutoff-regularized NJL/Gross--Neveu mean-field result in four spacetime dimensions.

For the first form factor,
\begin{equation}
f_I(\slashed\partial)=e^{-\slashed\partial/\Lambda},
\label{fI}
\end{equation}
one finds
\begin{equation}
f_I^2(\slashed p)=\cosh\left(\frac{2p}{\Lambda}\right)\Id-
\frac{1}{p}\sinh\left(\frac{2p}{\Lambda}\right)\slashed p,
\label{fI2}
\end{equation}
and hence
\begin{equation}
\Kcal_I(p;M,\Lambda)=
\frac{p^2\cosh(2p/\Lambda)+M^2}
{p^4+2M^2p^2\cosh(2p/\Lambda)+M^4} .
\label{KI}
\end{equation}
For the second form factor,
\begin{equation}
f_{II}(\slashed\partial)=e^{-i\slashed\partial/\Lambda},
\label{fII}
\end{equation}
the decomposition is
\begin{equation}
f_{II}^2(\slashed p)=\cos\left(\frac{2p}{\Lambda}\right)\Id-
\frac{i}{p}\sin\left(\frac{2p}{\Lambda}\right)\slashed p,
\label{fII2}
\end{equation}
which gives
\begin{equation}
\Kcal_{II}(p;M,\Lambda)=
\frac{p^2\cos(2p/\Lambda)+M^2}
{p^4+2M^2p^2\cos(2p/\Lambda)+M^4} .
\label{KII}
\end{equation}
Equations~\eqref{KI} and \eqref{KII} are the corrected kernels. They are obtained only after the matrix inversion has been carried out before the Dirac trace is taken. The exact nonlocal integrals generated by Eqs.~\eqref{KI} and \eqref{KII} are not elementary. We therefore follow the strategy used in the Dirac-like nonlocal spinor theory of Ref.~\cite{Nascimento:2025ngc}: introduce an intermediate scale $\Omega$ $M\ll \Omega\ll \Lambda\leq \Lambda_{\mathrm{UV}}$ and split the integral into infrared and ultraviolet pieces as follows
\begin{equation}
\Jcal_f=\Jcal_f^{\mathrm{IR}}+\Jcal_f^{\mathrm{UV}},
\label{eq:split}
\end{equation}
where the IR part is expanded in powers of $p/\Lambda$ and the UV part is expanded in powers of $M/p$. The matching scale $\Omega$ is not a new physical parameter; residual $\Omega$ dependence estimates the error of the truncated approximation. In the IR region the two kernels have the common expansion
\begin{equation}
\Kcal_\eta(p;M,\Lambda)=\frac{1}{p^2+M^2}
+\eta\,\frac{2p^4(p^2-M^2)}{\Lambda^2(p^2+M^2)^3}
+\mathcal{O}\left(\frac{p^4}{\Lambda^4}\right),
\label{eq:IRkernel}
\end{equation}
where $\eta=+1$ corresponds to $f_I$ and $\eta=-1$ corresponds to $f_{II}$. Therefore
\begin{equation}
\Jcal_\eta^{\mathrm{IR}}=\frac{1}{2}\left[\Omega^2-M^2\ln\left(1+\frac{\Omega^2}{M^2}\right)\right]
+\eta\frac{\mathcal{A}(M,\Omega)}{\Lambda^2}
+\mathcal{O}\left(\frac{\Omega^6}{\Lambda^4}\right),
\label{JIR}
\end{equation}
with
\begin{align}
\mathcal{A}(M,\Omega)=&\frac{1}{2(\Omega^2+M^2)^2}
\Bigg[\Omega^8-6M^2\Omega^6-27M^4\Omega^4-18M^6\Omega^2
\nonumber\\
&\hspace{2.5cm}
+18M^4(\Omega^2+M^2)^2\ln\left(1+\frac{\Omega^2}{M^2}\right)\Bigg].
\label{Afunc}
\end{align}
The first term in Eq.~\eqref{JIR} is the local contribution integrated only up to $\Omega$; the second term is the leading nonlocal correction. Its sign already shows the qualitative difference between the two entire functions: $f_I$ enhances the IR part of the gap integral, whereas $f_{II}$ suppresses it.

In the UV region, $p\gg M$, the kernels reduce to
\begin{equation}
\Kcal_I^{\mathrm{UV}}(p)=\frac{\cosh(2p/\Lambda)}{p^2}+\mathcal{O}\left(\frac{M^2}{p^4}\right),
\label{UVkernels}
\end{equation}
and
\begin{equation}
\Kcal_{II}^{\mathrm{UV}}(p)=\frac{\cos(2p/\Lambda)}{p^2}+\mathcal{O}\left(\frac{M^2}{p^4}\right)    
\end{equation}

The corresponding analytical integrals are
\begin{align}
\Jcal_I^{\mathrm{UV}}=&\left[
\frac{\Lambda p}{2}\sinh\left(\frac{2p}{\Lambda}\right)
-\frac{\Lambda^2}{4}\cosh\left(\frac{2p}{\Lambda}\right)
\right]_{\Omega}^{\Lambda_{\mathrm{UV}}}
+\mathcal{O}(M^2),
\label{JUV-I}\\
\Jcal_{II}^{\mathrm{UV}}=&\left[
\frac{\Lambda p}{2}\sin\left(\frac{2p}{\Lambda}\right)
+\frac{\Lambda^2}{4}\cos\left(\frac{2p}{\Lambda}\right)
\right]_{\Omega}^{\Lambda_{\mathrm{UV}}}
+\mathcal{O}(M^2).
\label{JUV-II}
\end{align}
Putting Eqs.~\eqref{JIR}, \eqref{JUV-I}, and \eqref{JUV-II} together gives the desired analytic gap equations:
\begin{equation}
\frac{M-m}{G}=\frac{M}{2\pi^2}\left[
\Jcal_I^{\mathrm{IR}}(M,\Omega,\Lambda)+\Jcal_I^{\mathrm{UV}}(\Omega,\Lambda,\Lambda_{\mathrm{UV}})
\right]
\label{analytic-gap-I}
\end{equation}
and
\begin{equation}
\frac{M-m}{G}=\frac{M}{2\pi^2}\left[
\Jcal_{II}^{\mathrm{IR}}(M,\Omega,\Lambda)+\Jcal_{II}^{\mathrm{UV}}(\Omega,\Lambda,\Lambda_{\mathrm{UV}})
\right].
\label{analytic-gap-II}
\end{equation}
These formulae are the gap-equation analog of the analytic IR/UV approximations used in Ref.~\cite{Nascimento:2025ngc}. They are particularly useful when $M\ll\Omega\ll\Lambda$ and when the UV cutoff is chosen of the same order as the nonlocality scale, $\Lambda_{\mathrm{UV}}\sim\Lambda$.

The local split approximation is recovered by replacing the UV kernels in Eq.~\eqref{UVkernels} by $1/p^2$, giving
\begin{equation}
\Jcal_{\mathrm{loc}}^{\rm split}=\frac{1}{2}\left[\Omega^2-M^2\ln\left(1+\frac{\Omega^2}{M^2}\right)\right]+\frac{1}{2}\left(\Lambda_{\mathrm{UV}}^2-\Omega^2\right)+\mathcal{O}(M^2/\Omega^2),
\label{Jlocsplit}
\end{equation}
which agrees with Eq.~\eqref{localJ} up to terms suppressed by $M^2/\Omega^2$ in the UV part.

Let us analyze the effect of nonlocality and comparison with the local case. In this case, the nonlocality scale modifies dynamical mass generation through the magnitude and sign of the gap integral. In the chiral limit $m=0$, the nontrivial solution satisfies
\begin{equation}
\frac{1}{G}=\frac{\Jcal_f(M,\Lambda,\Lambda_{\mathrm{UV}})}{2\pi^2}.
\label{chiral-gap}
\end{equation}
The critical coupling is obtained by taking $M\to0$,
\begin{equation}
G_c^{(f)}=\frac{2\pi^2}{\Jcal_f(0,\Lambda,\Lambda_{\mathrm{UV}})}.
\label{Gcdef}
\end{equation}
For the local theory,
\begin{equation}
G_c^{\mathrm{loc}}=\frac{4\pi^2}{\Lambda_{\mathrm{UV}}^2}.
\label{Gcloc}
\end{equation}
Using the $M	o0$ limit of the exact kernels and defining $s\equiv\Lambda_{\mathrm{UV}}/\Lambda$, the nonlocal critical couplings become
\begin{align}
G_c^{I}\Lambda^2&=\frac{2\pi^2}{\frac{s}{2}\sinh(2s)-\frac{1}{4}\left[\cosh(2s)-1\right]},
\label{GcI}\\
G_c^{II}\Lambda^2&=\frac{2\pi^2}{\frac{s}{2}\sin(2s)+\frac{1}{4}\left[\cos(2s)-1\right]} .
\label{GcII}
\end{align}
For the natural choice $\Lambda_{\mathrm{UV}}=\Lambda$ one finds
\begin{equation}
G_c^{\mathrm{loc}}\Lambda^2=4\pi^2\simeq39.48,
\quad
G_c^{I}\Lambda^2\simeq17.58,
\quad
G_c^{II}\Lambda^2\simeq196.19 .
\label{Gcnums}
\end{equation}
Thus the hyperbolic form factor strongly favors condensation, whereas the oscillatory form factor makes condensation harder. Physically, this occurs because $f_I$ enhances high-momentum contributions through $\cosh(2p/\Lambda)$, while $f_{II}$ introduces oscillatory cancellations through $\cos(2p/\Lambda)$.

This result gives a useful physical interpretation of the model. In the chiral limit, a nonzero solution of Eq.~\eqref{chiral-gap} appears only when the attractive interaction is strong enough to overcome the curvature of the effective potential at the origin. The values in Eq.~\eqref{Gcnums} show that this threshold is not universal once the fermion kinetic operator is made nonlocal. The hyperbolic kernel increases the available phase space for condensation, so the origin becomes unstable already at a smaller coupling. The oscillatory kernel produces cancellations in the same momentum range and therefore keeps the symmetric configuration stable up to much larger couplings. This provides a simple way to read the numerical results, where nonlocality can either catalyze or inhibit dynamical mass generation depending on the analytic form of the operator.

This interpretation parallels the usual NJL/GN picture of a condensate forming only when the attraction in the scalar channel is strong enough. In the local four-dimensional NJL model, the cutoff dependence of the critical coupling is a familiar consequence of treating the model as an effective theory. Here the same effective-theory logic remains, but the nonlocal form factor redistributes the momentum support of the gap integral before the cutoff is imposed. The comparison with the standard NJL reviews is therefore useful: it shows that the qualitative order-parameter language is conventional, while the quantitative shift of $G_c$ is a genuine effect of the nonlocal kinetic operator \cite{Klevansky:1992qe,Hatsuda:1994pi,Buballa:2003qv}.

The local limit is smooth in the infrared. For fixed $p,M$ and $\Lambda\to\infty$,
\begin{equation}
\Kcal_{I,II}(p;M,\Lambda)=\frac{1}{p^2+M^2}
\pm\frac{2p^4(p^2-M^2)}{\Lambda^2(p^2+M^2)^3}+\mathcal{O}(\Lambda^{-4}),
\label{smooth-local}
\end{equation}
where the upper sign corresponds to $f_I$ and the lower sign to $f_{II}$. This equation makes explicit that nonlocality is an irrelevant deformation at low momenta, but it becomes decisive once momenta probe the scale $p\sim\Lambda$. The comparison with the local model should therefore always state the UV prescription. If $\Lambda_{\mathrm{UV}}$ is taken much larger than $\Lambda$, the hyperbolic kernel grows rapidly and the model must be completed or renormalized with additional counterterms. If instead $\Lambda_{\mathrm{UV}}\sim\Lambda$, the model is interpreted as an effective nonlocal theory valid up to its intrinsic scale.

\section{Finite temperature and chemical potential}\label{s5}

In this section, we will examine the model at finite temperature.
The analysis is motivated by several physical questions that cannot be addressed at zero temperature. First, if the mass $M$ is dynamically generated, increasing the temperature is expected to disorder the condensate and may restore the symmetry that was broken by the zero-temperature solution. Second, the grand potential $\Omega(T,\mu;M)$ allows us to distinguish continuous crossovers from genuine phase transitions by comparing competing minima as functions of $T$ and $\mu$. Third, because the present model contains a nonlocal scale $\Lambda$, thermal fluctuations probe whether the restoration temperature and the phase-boundary structure are controlled only by the local NJL/Gross-Neveu mechanism or are substantially shifted by the analytic form factor. Thus, the purpose of this section is to prepare the framework for studying symmetry restoration, critical temperatures, and possible changes in the order of the transition induced by nonlocality.

A further motivation comes from the broader use of NJL-type four-fermion models in hot and dense matter. In QCD-inspired applications, scalar quark condensates describe chiral symmetry breaking and restoration, while at large chemical potential other four-fermion channels may support Cooper-pair condensates and color-superconducting phases \cite{Hatsuda:1994pi,Buballa:2003qv,Alford:2007xm}. The present work keeps only the scalar channel, so we do not claim to describe superconductivity in this minimal setup. Nevertheless, the same finite-$T$ and finite-$\mu$ machinery is precisely what would be needed to test whether a nonlocal Dirac operator enhances or suppresses such additional pairing channels in future extensions.

We will implement finite temperature in the nonlocal four-fermion model using the imaginary-time (Matsubara) formalism (see e.g. \cite{Matsubara:1955ws,Dolan:1973qd}; an excellent review on this approach is presented in \cite{Kapusta:2006pm}). To do so, we work in Euclidean spacetime with $\tau\in[0,\beta]$, $\beta\equiv 1/T$, and choose the standard antiperiodic boundary conditions for fermions, $\Psi(\tau+\beta,\mathbf{x})=-\Psi(\tau,\mathbf{x})$ (see e.g. \cite{Matsubara:1955ws}). At finite temperature and chemical potential $\mu$, the grand-canonical partition function at the homogeneous saddle is defined as follows
\begin{equation}
Z(T,\mu)=\int\!\mathcal{D}\bar\Psi\,\mathcal{D}\Psi\;
\exp\!\left\{-\int_0^\beta\!d\tau\int d^3x\;\left[\bar\Psi\Big(\slashed{\partial}_E\,f(\slashed{\partial}_E)-M-\mu\gamma_4\Big)\Psi+\frac{N}{2G}\bar\sigma^2\right]\right\},\label{eq60}
\end{equation}
where $\slashed{\partial}_E\equiv \gamma_4\partial_\tau+\gamma_i\partial_i$ (Euclidean), and $\gamma_4$ is the Euclidean time gamma matrix. The sign of the auxiliary-field term is positive in the Euclidean action, so that the classical contribution to the thermodynamic potential is positive and agrees with the zero-temperature potential. By integrating out the fermions, one obtains the full mean-field grand-potential density $\Omega_{\rm full}=N\Omega$, where the potential per fermion flavor is
\begin{equation}
\Omega(T,\mu;M)=\frac{(M-m)^2}{2G}
- T \sum_{n\in\mathbb{Z}} \int\!\frac{d^3p}{(2\pi)^3}\;
\ln{\rm det}_D\!\Big[\,i\slashed{P}_n\,f(i\slashed{P}_n)-M\,\Big].
\label{eq61}
\end{equation}
where ${\rm det}_D$ is over Dirac indices. Besides, in the Matsubara formalism, one has
\begin{align}
\omega_n&=(2n+1)\pi T,\label{eq62}
\\ P_n^\mu &= \big(\omega_n-i\mu,\;\mathbf{p}\big),\label{eq63}
\\ P_n^2 &= (\omega_n-i\mu)^2+\mathbf{p}^2.
\label{eq64}
\end{align}
With these definitions, the mean-field condition $\partial\Omega/\partial M=0$ gives the following gap equation at finite temperature
\begin{equation}
\frac{M-m}{G}=4M T\sum_{n\in\mathbb Z}\int\!\frac{d^3p}{(2\pi)^3}\,
\Kcal_f(P_n;M,\Lambda),
\label{thermal-gap}
\end{equation}
with the same matrix-inverted kernels as at zero temperature, now evaluated at $P=P_n$.

Equation~\eqref{thermal-gap} is the finite-temperature counterpart of the zero-temperature mean-field condition. Its solutions $M(T,\mu)$ determine the thermal evolution of the condensate. In the chiral limit, a solution that decreases continuously to zero as $T$ increases indicates second-order symmetry restoration, whereas the coexistence of separated local minima in $\Omega(T,\mu;M)$ would indicate a first-order transition. For nonzero bare mass $m$, the sharp transition may be replaced by a crossover, but the temperature dependence of $M$ still measures how efficiently thermal excitations melt the dynamically generated mass.

The contour representation is cleanest if the complex variable is chosen as $z=i\omega_n$, and the Fermi function is
\begin{equation}
n_F(z)=\frac{1}{e^{\beta z}+1} .
\label{nF-clean}
\end{equation}
For a meromorphic function $F(z)$, after the usual zero-temperature subtraction when needed,
\begin{equation}
T\sum_n F(i\omega_n)=\sum_{z_\star\in\text{\{poles of }F\}} n_F(z_\star)\,\mathrm{Res}_{z=z_\star}F(z).
\label{contour-clean}
\end{equation}
With the convention $z=i\omega_n$ used here, the residue formula carries the plus sign shown in Eq.~\eqref{contour-clean}. This sign is fixed by the local benchmark below.
In the local case,
\begin{equation}
F_{\rm loc}(z)=\frac{1}{E_{\mathbf p}^2-(z+\mu)^2},
\quad E_{\mathbf p}=\sqrt{\mathbf p^2+M^2},
\label{Floc-clean}
\end{equation}
with poles $z=-\mu\pm E_{\mathbf p}$. Equation~\eqref{contour-clean}, together with the zero-temperature contribution, gives
\begin{equation}
T\sum_n\frac{1}{(\omega_n-i\mu)^2+E_{\mathbf p}^2}
=\frac{1}{2E_{\mathbf p}}\left[1-n_F(E_{\mathbf p}-\mu)-n_F(E_{\mathbf p}+\mu)\right],
\label{standard-sum}
\end{equation}
and hence
\begin{equation}
\frac{M-m}{G}=4M\int\!\frac{d^3p}{(2\pi)^3}\frac{1}{2E_{\mathbf p}}
\left[1-n_F(E_{\mathbf p}-\mu)-n_F(E_{\mathbf p}+\mu)\right]
\label{local-thermal-gap}
\end{equation}
in the local limit.

The local expression in Eq.~\eqref{local-thermal-gap} is an important benchmark. The factor $1-n_F(E_{\mathbf p}-\mu)-n_F(E_{\mathbf p}+\mu)$ has a transparent interpretation: thermal particles and antiparticles reduce the available phase space for pairing and therefore weaken the condensate. The nonlocal theory must reproduce this structure when $\Lambda\rightarrow\infty$; deviations from it at finite $\Lambda$ quantify how the nonlocal operator changes the quasiparticle spectrum, the residues of the poles, and ultimately the critical temperature associated with symmetry restoration.

For the nonlocal theory, the same contour method gives a quasiparticle representation. Let
\begin{equation}
F_f(z;\mathbf p)=\Kcal_f\left(P(z);M,\Lambda\right),
\quad
P^2(z)=\mathbf p^2-(z+\mu)^2 .
\label{Ff}
\end{equation}
If the scalar denominator has isolated simple zeros at $z=E_\ell(\mathbf p)$, then
\begin{equation}
T\sum_n F_f(i\omega_n;\mathbf p)=
\sum_\ell n_F(E_\ell)\frac{N_f(E_\ell;\mathbf p)}{\partial_zD_f(E_\ell;\mathbf p)}
+\text{complex-pole/cut contributions},
\label{nonlocal-contour}
\end{equation}
where $F_f=N_f/D_f$. When physical poles occur in particle-antiparticle pairs, Eq.~\eqref{nonlocal-contour} reorganizes into the familiar thermal occupation structure, but with nonlocal residues and nonlocal quasiparticle energies. The standard factor $1-n_F(E-\mu)-n_F(E+\mu)$ is recovered when only the local branch survives.

From the point of view of phase structure, Eq.~\eqref{nonlocal-contour} is the key difference between the present model and the local theory. The thermal occupation factors are still controlled by poles in the complex-energy plane, but the positions and residues of those poles are now functions of $\Lambda$ and of the chosen entire operator. Consequently, a change of form factor may shift the restoration temperature, alter the curvature of the grand potential near $M=0$, or even modify the order of the transition if additional nonlocal branches become relevant. A complete numerical phase diagram is beyond the scope of the present derivation, but the formulae above identify exactly which ingredients must be evaluated to obtain it.

Let us analyze the pole structure and stability. In this case, the pole structure is a necessary consistency check because the form factor is a function of the Dirac operator. Setting $\mu=0$ and analytically continuing to time-like momenta, define
$q=\sqrt{E^2-\mathbf p^2}$. The pole equations associated with Eqs.~\eqref{KI} and \eqref{KII} are
\begin{align}
D_I(q)&=q^4-2M^2q^2\cos\left(\frac{2q}{\Lambda}\right)+M^4=0,
\label{pole-I}\\
D_{II}(q)&=q^4-2M^2q^2\cosh\left(\frac{2q}{\Lambda}\right)+M^4=0.
\label{pole-II}
\end{align}
Both reduce to $(q^2-M^2)^2=0$ as $\Lambda\to\infty$. For $M/\Lambda\ll1$, the physical solution is continuously connected to $q=M$, so the local quasiparticle branch is preserved. However, Eqs.~\eqref{pole-I} and \eqref{pole-II} can also admit additional real or complex solutions depending on $M/\Lambda$ and on the analytic continuation chosen for the form factor. These extra solutions must not be ignored: real extra poles would enlarge the spectrum, while complex poles would signal Lee-Wick-type excitations or an instability unless a consistent contour prescription is supplied.

For this reason, the effective theory should be used in the conservative regime
\begin{equation}
M\ll \Lambda,
\quad
p\lesssim\Lambda_{\mathrm{UV}}\sim\Lambda,
\label{stability-window}
\end{equation}
where the physical pole is the local branch deformed by small nonlocal corrections and the IR expansion \eqref{smooth-local} is reliable. A full spectral analysis beyond this window is possible, but it requires specifying the nonlocal completion and the pole prescription. This point is especially important for the oscillatory form factor, whose denominator may become small near trigonometric zeros when the cutoff is extended too far above $\Lambda$.

These observations strengthen the physical meaning of the finite-temperature analysis. The conservative window in Eq.~\eqref{stability-window} is the regime in which the thermal problem can be interpreted as a deformation of the familiar local scenario: the condensate melts because of thermal occupation, while nonlocality changes the rate of melting through the kernel and the quasiparticle residues. Outside this window, additional poles may compete with the local branch, and the phase structure can no longer be inferred from the standard NJL intuition alone. This is precisely why the pole analysis and the finite-temperature gap equation must be studied together.

The analytic results above are complemented by simple numerical illustrations. In all plots we set $\Lambda_{\mathrm{UV}}=\Lambda$ and use dimensionless variables $x=p/\Lambda$, $\mu_M=M/\Lambda$, and $g=G\Lambda^2$.

\begin{figure}[H]
\centering
\includegraphics[width=0.74\linewidth]{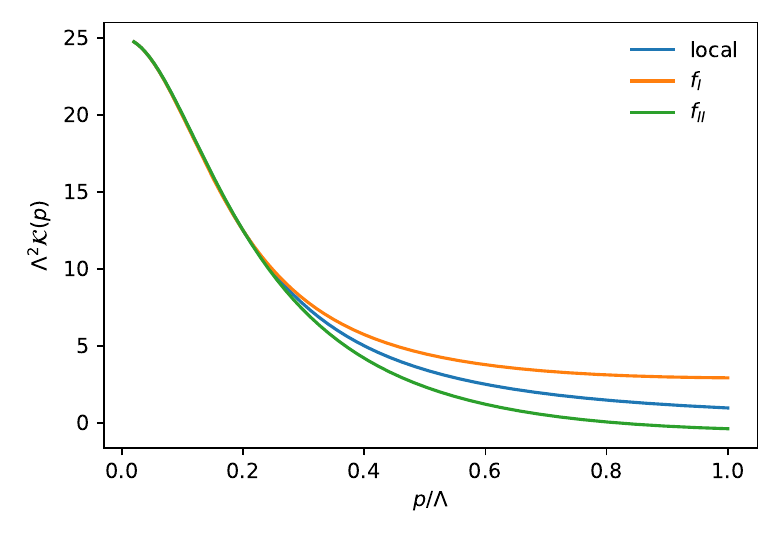}
\caption{Dimensionless kernels $\Lambda^2\Kcal(p)$ for $M/\Lambda=0.2$. The hyperbolic form factor enhances the kernel at moderate and large momenta, while the oscillatory form factor suppresses it relative to the local case.}
\label{fig:kernel}
\end{figure}

\begin{figure}[H]
\centering
\includegraphics[width=0.74\linewidth]{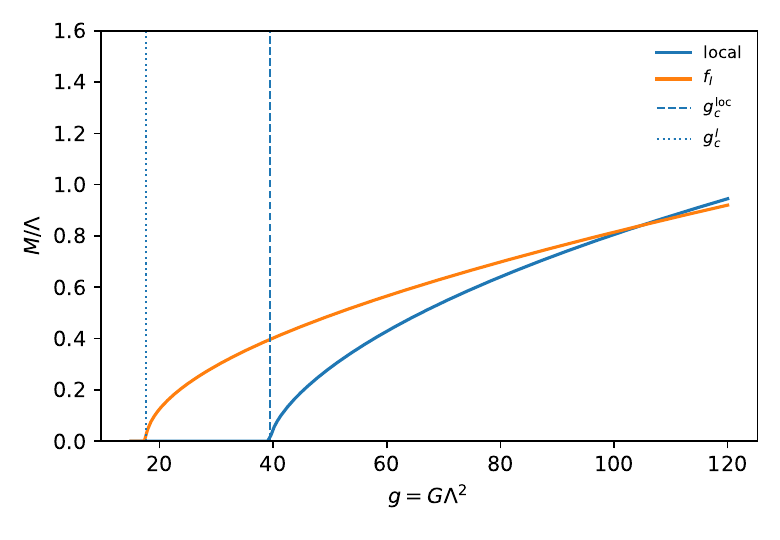}
\caption{Dynamical mass in the chiral limit as a function of $g=G\Lambda^2$ for the local theory and for $f_I$. The lower critical coupling for $f_I$ reflects the enhancement of the gap integral by the hyperbolic nonlocal kernel.}
\label{fig:massg}
\end{figure}

\begin{figure}[H]
\centering
\includegraphics[width=0.74\linewidth]{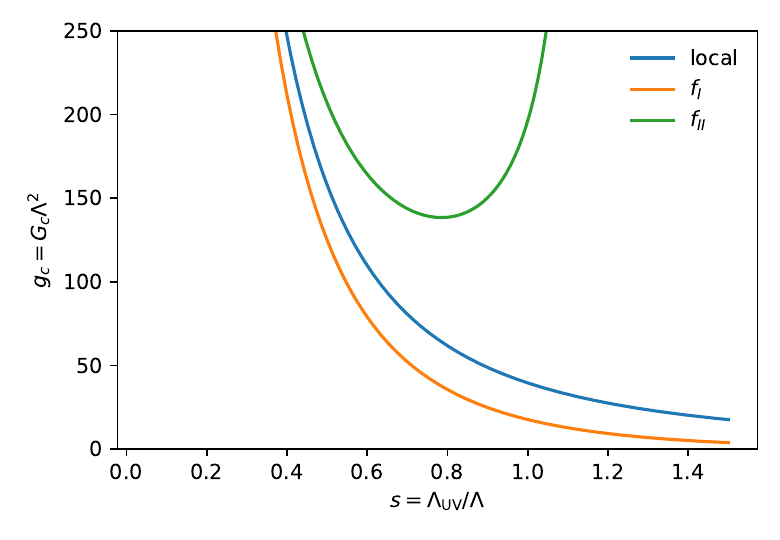}
\caption{Analytic estimation of the critical coupling as a function of $s=\Lambda_{\mathrm{UV}}/\Lambda$. The hyperbolic form factor lowers $g_c$, whereas the oscillatory form factor raises it for $s\lesssim1$.}
\label{fig:critical}
\end{figure}

\begin{figure}[H]
\centering
\includegraphics[width=0.74\linewidth]{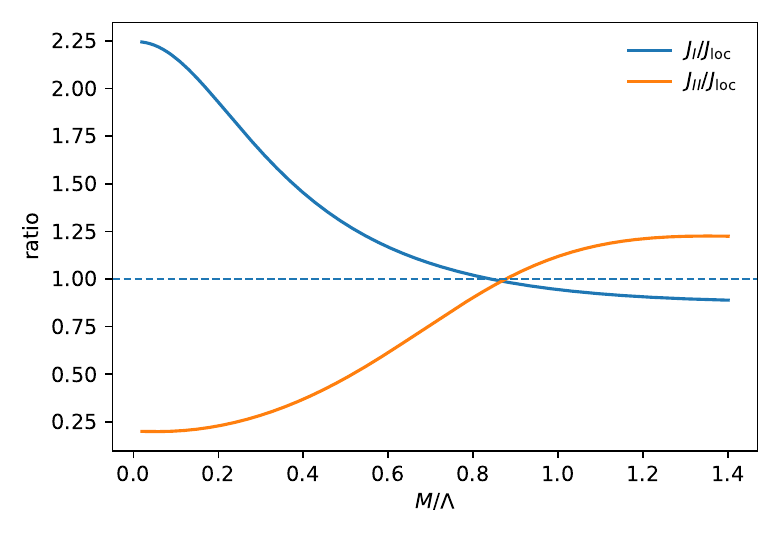}
\caption{Ratio between the exact cutoff gap integrals and the local integral for $\Lambda_{\mathrm{UV}}=\Lambda$. The sign and magnitude of the nonlocal effect depend strongly on the analytic form of the  Dirac-like form factor.}
\label{fig:ratio}
\end{figure}

The figures confirm the analytic discussion. The model with $f_I$ develops a nonzero mass at smaller coupling than the local theory. Conversely, $f_{II}$ is dominated by cancellations at the scale $p\sim\Lambda$, and therefore the interaction must be stronger to trigger condensation. These results also show why the UV prescription cannot be separated from the physical interpretation of the nonlocal model.

More broadly, the discussion shows that the role of nonlocality is not limited to improving or regulating ultraviolet behavior. In this model it actively participates in the mechanism of dynamical mass generation by changing the balance between the classical auxiliary-field energy and the fermionic determinant. This balance is what determines the vacuum at zero temperature and the possible restoration pattern at finite temperature. Therefore, the form factor should be regarded as a physical ingredient of the effective theory, not only as a computational device.

\section{Final remarks}\label{s6}

In this paper, we have developed a nonlocal extension of the four-fermion model by combining a Dirac kinetic operator modified by an entire form factor with a local quartic fermionic interaction. This construction offers a natural way to investigate how nonlocality can affect dynamical mass generation while preserving the familiar structure of fermionic self-interactions. Our main result was the derivation of the effective potential and of the corresponding gap equation in a framework where the nonlocality is encoded in the fermion propagator. A crucial aspect of the calculation is that the form factor $f(\slashed{p})$ must be handled as a matrix-valued function. Once this point is treated consistently, the gap equation can be written in terms of a scalar kernel that makes the effect of nonlocality transparent. In this form, one sees clearly how the scale $\Lambda$ modifies the mean-field dynamics and how the local NJL-type expression is recovered when $\Lambda \to \infty$.

We have also obtained explicit kernels for two representative choices of nonlocal operator, namely $f_I(\slashed{\partial})=e^{-\slashed{\partial}/\Lambda}$ and $f_{II}(\slashed{\partial})=e^{-i\slashed{\partial}/\Lambda}$. For both cases, the renormalized gap equation was written in a form suitable for practical analysis, separating the standard local contribution from a finite nonlocal remainder. This makes the comparison with the usual four-fermion theory especially clear and provides a useful starting point for numerical investigations of the dynamical mass as a function of the nonlocality scale. With the IR/UV split used in the Dirac-like nonlocal spinor theory, the resulting formulae, Eqs.~\eqref{analytic-gap-I} and \eqref{analytic-gap-II}, make the effect of nonlocality transparent. The hyperbolic form factor enhances the gap integral and lowers the critical coupling, while the oscillatory form factor suppresses the integral and raises the critical coupling. This difference is not a small technical detail; it shows that different entire functions of $\slashed\partial$ can lead to qualitatively distinct patterns of dynamical mass generation.

At finite temperature and chemical potential, the Matsubara formalism allowed us to extend our analysis in a natural way, Within this framework, the resulting thermal gap equations show that the nonlocal modifications survive in the quasiparticle sector through the analytic structure of the kernel, and the contour representation makes explicit how the thermal behavior is governed by the pole structure of the nonlocal propagator. Once again, the local limit reproduces the standard NJL/Gross-Neveu thermal relation, which serves as an important consistency check of the whole construction.

Therefore, these results establish a interesting framework for studying spontaneous symmetry breaking and thermal restoration in nonlocal fermionic theories. Although the present work is mainly focused on the formal derivation of the model and its gap equations, it already sets the stage for several natural developments, such as a numerical study of the phase diagram, a comparison between different classes of entire form factors, and a more detailed analysis of the associated quasiparticle spectrum. In this sense, the model discussed here may serve as a useful laboratory for understanding how nonlocality reshapes familiar mechanisms of dynamical mass generation in fermionic quantum field theory.

It is important to highlight that several directions deserve further investigation as natural continuations of this work. A first step is compare different classes of entire form factors in order to understand which features of dynamical mass generation are genuinely universal and which depend on the specific analytic choice for the nonlocal operator. Another promising direction is the study of the quasiparticle spectrum, especially the role of possible complex poles or branch-cut contributions and their implications for stability, causality, and thermodynamic behavior. One may also extend the model to chirally symmetric, vector, or axial interaction channels, and investigate whether nonlocality modifies the usual pattern of spontaneous symmetry breaking and restoration. Finally, the formalism developed here could serve as a starting point for exploring $1/N$ corrections, transport properties, or possible applications to effective descriptions of strongly correlated fermionic systems in which nonlocal effects are expected to play a relevant role.

\section*{Acknowledgments}
The authors thank the Coordena\c{c}\~ao de Aperfei\c{c}oamento de Pessoal de N\'ivel Superior (CAPES), the Paraiba State Research Foundation (FAPESQ-PB), and the Conselho Nacional de Desenvolvimento Cient\'ifico e Tecnol\'ogico (CNPq). Fernando M. Belchior thanks CNPq for grant No. 151845/2025-5. Albert Yu. Petrov thanks FAPESQ-PB (process No. 150891/2023-7) and CNPq (grant No. 303777/2023-0). Paulo J. Porf\'irio thanks FAPESQ-PB (process No. 150891/2023-7) and CNPq (grant No. 307628/2022-1).


\end{document}